\documentclass[pra,aps,preprint, floats,showpacs,preprintnumbers]{revtex4}

\usepackage[]{graphicx}
\usepackage{amsmath}
\usepackage{amssymb}

\usepackage{epsfig}
\usepackage{wrapfig}
\usepackage{color}

\def\##1{{\bf #1}}
\def\=#1{\underline{\underline #1}}

\def\.{\mbox{ \tiny{$^\bullet$} }}

\def\epso{\epsilon_{0}}
\def\lambdao{\lambda_{ 0}}
\def\muo{\mu_{ 0}}
\def\ko{k_{ 0}}
\def\etao{\eta_{0}}

\def\eps{\epsilon}

\def\epsmet{\eps_{met}}

\def\sp{\mathbf s}
\def\pinc{{\mathbf p}^+}
\def\pref{{\mathbf p}^-}

\def\Einc{{\mathbf E}_{inc}({\bf r})}
\def\Erefl{{\mathbf E}_{ref}({\bf r})}
\def\Etr{{\mathbf E}_{tr}({\bf r})}

\def\Hinc{{\mathbf H}_{inc}({\bf r})}
\def\Hrefl{{\mathbf H}_{ref}({\bf r})}
\def\Htr{{\mathbf H}_{tr}({\bf r})}

\def\pn{^{(n)}}
\def\pom{^{(m)}}
\def\bE{{\bf E}}
\def\bH{{\bf H}}
\def\br{{\bf r}}

\def\ux{\hat{\#u}_x}
\def\uy{\hat{\#u}_y}
\def\uz{\hat{\#u}_z}

\begin{document}

\title{Grating-coupled excitation of multiple surface plasmon-polariton waves}

\author{Muhammad Faryad}
\author{Akhlesh Lakhtakia\footnote{Corresponding author}}\email{akhlesh@psu.edu}

\affiliation{Nanoengineered Metamaterials Group (NanoMM), Department of Engineering Science and
Mechanics, 
Pennsylvania State University, University Park, PA  16802-6812, USA}

\begin{abstract}
The excitation of multiple surface-plasmon-polariton~(SPP) waves of different linear polarization states and phase speeds by a surface-relief grating formed by a metal and a rugate filter, both of finite thickness, was studied theoretically, using rigorous coupled-wave-analysis. The incident plane wave can be either $p$  or $s$ polarized. The excitation of SPP waves is indicated by the presence of those peaks in the plots of absorbance vs. the incidence angle that are independent of the thickness of the rugate filter.  The absorbance peaks representing the excitation of $s$-polarized SPP waves are narrower than those representing $p$-polarized SPP waves. Two incident plane waves propagating in different directions may excite the same SPP wave. A line source could excite several SPP waves simultaneously.
\end{abstract}


\maketitle
\section{Introduction}
Surface plasmon-polariton~(SPP) waves are surface waves guided by a planar interface of a metal and a dielectric material. SPP waves find  
applications for sensing, imaging and communication \cite{Maier07,Homola_book}.
If the dielectric partnering material is isotropic and homogeneous, only one SPP wave---that too, of the $p$-polarization state---can be guided by the metal-dielectric interface at a given frequency ~\cite{ZSM,Dragoman08}.  
If a periodic nonhomogeneity normal to the wave-guiding interface is introduced in the  dielectric partnering material, multiple SPP waves with different polarization states, phase speeds, and spatial profiles can be guided by the metal-dielectric interface. This has recently been shown both theoretically~\cite{Polo2010,PL2009,PartIV} and experimentally \cite{DPL,Gilani,PartIII}.  In all of these studies, the dielectric partnering material is also locally orthorhombic.

Very recently, we have solved a canonical boundary-value problem~\cite{Faryadjosab} to show that multiple SPP waves can be guided even if the dielectric partnering material is isotropic---provided that material is also
periodically nonhomogeneous normal to the interface.  This is a very attractive result, because both partnering materials are isotropic and because the dielectric partnering material can be fabricated as a
rugate filter \cite{Bovard,Overend1992,Lorenzo2005,Lee2006,Lakh2011}.

The canonical boundary-value problem does not possess direct practical significance, because both partnering materials are assumed to be semi-infinite normal to the planar interface. Therefore, we set out to investigate the excitation of multiple SPP waves by the periodically corrugated interface of a metal and a rugate filter. This grating-coupled configuration \cite[pp.~35--41]{Homola_book} is   popular, when the dielectric partnering material is homogeneous, because
it allows the excitation of an SPP wave by a nonspecular Floquet harmonic.
The interplay of the periodic nonhomogeneity of the dielectric partnering material and a periodically corrugated interface is phenomenologically rich \cite{McCall2003,McIlroy2005}, and should lead to the excitation of multiple SPP waves as different Floquet harmonics.

 The relevant boundary-value problem was formulated using the  rigorous coupled-wave analysis~(RCWA)~\cite{Moharam82,Li93}. In this numerical technique, the constitutive parameters are expanded in terms of Fourier series with known
 expansion coefficients, and
 the electromagnetic field phasors  are expanded in terms of Floquet harmonics whose coefficients are determined by substitution in the frequency-domain Maxwell curl postulates.  The accuracy of solution is conventionally held to depend only on the number of Floquet harmonics actually used in the computations~\cite{Moharam95}.  The RCWA has been used to solve for scattering by  a variety of surface-relief gratings~\cite{Li93,Moharam95,Glytsis,Wang04}, generally with both partnering materials being homogeneous.

The theoretical formulation of the   boundary-value problem is provided in Sec.~\ref{theory} and the numerical results are discussed in Sec.~\ref{nrd}. Concluding remarks are presented in Sec.~\ref{conc}.
An $\exp(-i\omega t)$ time-dependence is implicit, with $\omega$
denoting the angular frequency. The free-space wavenumber, the
free-space wavelength, and the intrinsic impedance of free space are denoted by $\ko=\omega\sqrt{\epso\muo}$,
$\lambdao=2\pi/\ko$, and
$\etao=\sqrt{\muo/\epso}$, respectively, with $\muo$ and $\epso$ being  the permeability and permittivity of
free space. Vectors are in boldface, 
column vectors are in boldface and enclosed within square brackets, and
matrixes are underlined twice and square-bracketed. The Cartesian unit vectors are
identified as $\ux$, $\uy$, and $\uz$. The superscript T denotes the transpose.

\section{Boundary-value problem}\label{theory}
\subsection{Description}
\begin{figure}[!ht]
\begin{center}
\includegraphics[width=4.5in]{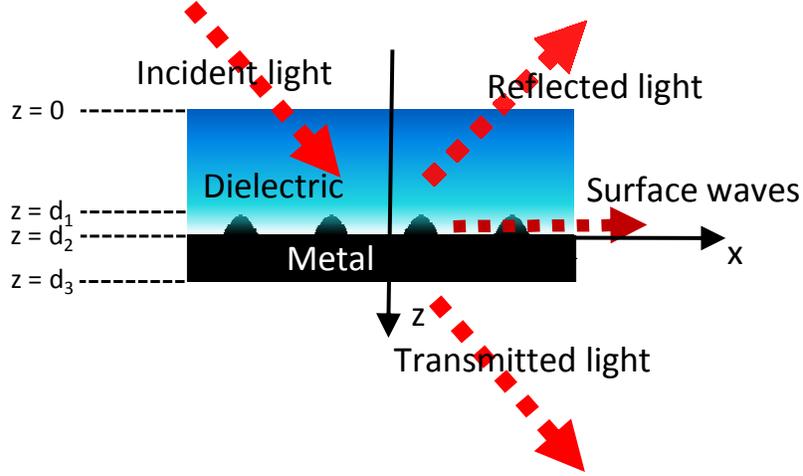}
\caption{Schematic of the boundary-value problem solved using the RCWA.}
\label{geom}
\end{center}
\end{figure}

Let us consider the schematic of the boundary-value problem shown in Fig.~\ref{geom}. The regions $z<0$ and $z>d_3$ are vacuous, the region $0\leq z\leq d_1$ is occupied by the dielectric partnering material with relative permittivity $\epsilon_d(z)$, and the region $d_2\leq z\leq d_3$ by the metallic partnering material with spatially uniform relative permittivity $\epsilon_m$. The region $d_1< z<d_2$ contains a surface-relief grating of period $L$ along the $x$ axis. The relative permittivity $\epsilon_g(x,z)=\epsilon_g(x\pm{L},z)$ in this region  is taken to be as  
 \begin{eqnarray}
\epsilon_g(x,z)=
\begin{cases}
{(1/2)}{\left[\epsilon_m+\epsilon_d(z)\right]}-[\epsilon_m-\epsilon_d(z)]\\
\quad \times\left\{{\cal U}\left[d_2-z-g(x)\right]-{1\over 2}\right\}\,,& x\in(0,L_1)\,,\\
\epsilon_d(z)\,,& x\in(L_1,L)\,,
\end{cases} 
\label{grating}
\end{eqnarray}
for $z\in\left(d_1,d_2\right)$, with
\begin{equation}
g(x)=(d_2-d_1)\sin \left(\frac{\pi x}{L_1}\right)\,,\qquad L_1\in(0,L)\,,
\label{ourgrating}
\end{equation}
and
\begin{equation}
{\cal U}(\zeta)=
\begin{cases}
&1\,,\qquad \zeta\ge 0\,,\\
&0\,,\qquad \zeta<0\,.
\end{cases}
\end{equation}
The depth of the surface-relief grating defined by Eq.~(\ref{ourgrating}) is $d_2-d_1$. This particular grating shape is chosen for the ease of fabrication; however, the theoretical formulation   given in the remainder of this section is independent of the shape of the surface-relief grating.


In the vacuous half-space $z\leq 0$,
let a plane wave propagating in the $xz$ plane at an angle $\theta$ to the $z$ axis, be incident on the
structure. Hence, the incident, reflected, and transmitted field phasors may be written in terms of Floquet harmonics as follows:
\begin{eqnarray}
\Einc&=&{\sum_{n\in \mathbb{Z}}}\left(\sp_n a_s^{(n)}+\pinc_na_p^{(n)}\right)
\exp\left[i\left(k_x^{(n)}x+k_z^{(n)}z\right)\right]\,, z\leq0\,,\label{einc}\\[5pt]
\Hinc&=& {\etao}^{-1}\, \sum_{n\in \mathbb{Z}}\left( \pinc_n a_s^{(n)}-\sp_na_p^{(n)}\right)
\exp\left[i\left(k_x^{(n)}x+k_z^{(n)}z\right)\right]\,, z\leq0\,,\label{hinc}\\[5pt]
\Erefl&=&\sum_{n\in \mathbb{Z}}\left( \sp_n r_s^{(n)}+\pref_nr_p^{(n)}\right)
\exp\left[i\left(k_x^{(n)}x-k_z^{(n)}z\right)\right]\,, z\leq0\,,\label{erefl}\\[5pt]
\Hrefl&=& {\etao}^{-1}\,  \sum_{n\in \mathbb{Z}}\left( \pref_n r_s^{(n)}-\sp_n r_p^{(n)}\right)
\exp\left[i\left(k_x^{(n)}x-k_z^{(n)}z\right)\right]\,, z\leq0\,,\label{hrefl}\\[5pt]
\Etr&=&\sum_{n\in \mathbb{Z}}\left(\sp_n t_s^{(n)}+\pinc_nt_p^{(n)}\right)
\exp\left\{i\left[k_x^{(n)}x+k_z^{(n)}\left(z-d_3\right)\right]\right\}\,, z\ge d_3\,,\label{etr}\\[5pt]
\Htr&=&{\etao}^{-1}\, \sum_{n\in \mathbb{Z}}\left(\pinc_n t_s^{(n)}-\sp_nt_p^{(n)}\right)
\exp\left\{i\left[k_x^{(n)}x+k_z^{(n)}\left(z-d_3\right)\right]\right\}\,,  z\ge d_3\,,\label{htr}
\end{eqnarray}
where  $k_x\pn=\ko\sin\theta+n\kappa_x$, $\kappa_x=2\pi/L$, and
\begin{equation}
k_z^{(n)}=
\begin{cases}
+\sqrt{\ko^2-{(k_x^{(n)})}^2}\,,& \ko^2> {(k_x^{(n)})}^2\\[5pt]
+i\sqrt{{(k_x^{(n)})}^2-\ko^2}\,,& \ko^2< {(k_x^{(n)})}^2
\end{cases}\,.\label{kz}
\end{equation}
The unit vectors 
\begin{equation}
\sp_n=\uy
\end{equation}
and
\begin{equation}
p_n^\pm=\mp{k_z^{(n)}\over\ko}\ux+{k_x^{(n)}\over \ko}\uz
\label{pn}
\end{equation}
represent the $s$- and $p$-polarization states, respectively. 

\subsection{Coupled ordinary differential equations}

The relative permittivity in the region $0\leq z\leq d_3$ can be expanded as a Fourier series with respect to $x$, viz.,
\begin{equation}
\epsilon(x,z)=\sum_{n\in \mathbb{Z}} \epsilon^{(n)}(z)\exp (i n \kappa_x x)\,, \qquad z\in\left[0,d_3\right]\,,
\label{perm}
\end{equation}
where $\kappa_x=2\pi/L$, 
\begin{equation}
\epsilon^{(0)}(z)=
\begin{cases}
\epsilon_d(z)\,,& z\in\left[0,d_1\right]\,,\\
{1\over L}\int_0^L\epsilon_g(x,z) d x\,,& z\in\left(d_1,d_2\right)\,,\\
\epsilon_m\,,& z\in\left[d_2,d_3\right]\,,
\end{cases}\label{perm0}
\end{equation}
and
\begin{equation}
\epsilon^{(n)}(z)=
\begin{cases}
{1\over L}\int_0^L\epsilon_g(x,z) \exp(-in\kappa_xx)d x\,,& z\in\left[d_1,d_2\right]\\
0\,,& \rm{otherwise}
\end{cases}\,;\forall n\neq 0\,.
\label{permn}
\end{equation}
The field phasors may be written in the region $0\leq z\leq d_3$ in terms of Floquet harmonics as
\begin{equation}
\left.\begin{array}{l}
\bE(\br)=\displaystyle{\sum_{n\in \mathbb{Z}}} \left[E_x\pn(z)\ux+E_y\pn(z)\uy+E_z\pn(z)\uz\right]\exp(ik_x\pn x) \\[5pt]
\bH(\br)=\displaystyle{\sum_{n\in \mathbb{Z}}}\left[H_x\pn(z)\ux+H_y\pn(z)\uy+H_z\pn(z)\uz\right]\exp(ik_x\pn x) 
\end{array}
\right\}\,,\quad
z\in\left[0,d_3\right]\,,\label{field}
\end{equation}
with unknown functions $E_{x,y,z}\pn(z)$ and $H_{x,y,z}\pn(z)$.

Substitution of Eqs.~(\ref{perm}) and (\ref{field}) in the frequency-domain Maxwell curl postulates results in a system of four ordinary differential equations and two algebraic equations as follows:
\begin{eqnarray}
&&\frac{d}{dz}E_x\pn(z)-ik_x\pn E_z\pn(z)=i\ko{\etao} H_y\pn(z)\,,\label{max1}\\
&&\frac{d}{dz}E_y\pn(z)=-i\ko{\etao} H_x\pn(z)\,,\label{max2}\\
&&k_x\pn E_y\pn(z)=\ko{\etao} H_z\pn(z)\,,\label{max3}\\
&&\frac{d}{dz}H_x\pn(z)-ik_x\pn H_z\pn(z)=-\frac{i\ko}{{\etao}}\sum_{m\in \mathbb{Z}}\epsilon^{(n-m)}(z)E_y\pom(z)\,,\label{max4}\\
&&\frac{d}{dz}H_y\pn(z)=\frac{i\ko}{{\etao}}\sum_{m\in \mathbb{Z}}\epsilon^{(n-m)}(z)E_x\pom(z)\,,\label{max5}\\
&&k_x\pn H_y\pn(z)=-\frac{\ko}{{\etao}}\sum_{m\in \mathbb{Z}}\epsilon^{(n-m)}(z)E_z\pom(z)\,.\label{max6}
\end{eqnarray}
Equations~(\ref{max1})--(\ref{max6}) hold $\forall{z}\in\left(0,d_3\right)$ and ${\forall}n\in \mathbb{Z}$. These equations can be
reformulated into an infinite system of coupled first-order ordinary differential equations. This system can not be implemented on a digital computer. Therefore, we restrict $|n|\leq N_t$ and then define the column $(2N_t+1)$-vectors
 \begin{equation}
 [\#X_\sigma(z)]=[X_\sigma^{(-N_t)}(z),~X_\sigma^{(-N_t)}(z),~...,~X_\sigma^{(0)}(z),~...,~X_\sigma^{(N_t-1)}(z),~X_\sigma^{(N_t)}(z)]^T\,,
 \end{equation}
 for $X\in\left\{E,H\right\}$ and $\sigma\in\left\{x,y,z\right\}$. Similarly, we define $(2N_t+1)\times(2N_t+1)$-matrixes
\begin{eqnarray}
[\=K_x]={\rm{diag}}[k_x\pn]\,,\qquad
[\=\epsilon(z)]=\left[\epsilon^{(n-m)}(z)\right]\,,
\end{eqnarray}
where ${\rm{diag}}[k_x\pn]$ is a diagonal matrix.

Substitution of Eqs.~(\ref{max3}) and (\ref{max6}) into (\ref{max1}), (\ref{max2}), (\ref{max4}) and (\ref{max5}), to eliminate $E_z\pn$ and $H_z\pn\, \forall {n \in \mathbb{Z}}$, gives the matrix ordinary differential equation
\begin{equation}
\frac{d}{dz}\left[\#f(z)\right]=i\left[\=P(z)\right]\cdot\left[\#f(z)\right]\,,\qquad z\in\left(0,d_3\right)\,,\label{mode}
\end{equation}
where the column vector $[\#f(z)]$ with $4(2N_t+1)$ components is defined as 
\begin{equation}
\left[\#f(z)\right]=\left[\left[\#E_x(z)\right]^T,~~\left[\#E_y(z)\right]^T,~~\etao\left[\#H_x(z)\right]^T,~~\etao\left[\#H_y(z)\right]^T\right]^T
\end{equation}
and the $4(2N_t+1)\times4(2N_t+1)$-matrix $\left[\=P(z)\right]$ is given by
\begin{equation}
\left[\=P(z)\right]=\left[
\begin{array}{cccc}
\left[\=0\right] & \left[\=0\right] & \left[\=0\right]&\left[\=P_{14}(z)\right]\\
\left[\=0\right] & \left[\=0\right] & -\ko\left[\=I\right] & \left[\=0\right]\\
\left[\=0\right] & \left[\=P_{32}(z)\right] & \left[\=0\right] & \left[\=0\right]\\
\left[\=P_{41}(z)\right] & \left[\=0\right] & \left[\=0\right] & \left[\=0\right]
\end{array}
\right]\,.\label{Pmat}
\end{equation}
Whereas $\left[\=0\right]$ is the $(2N_t+1)\times(2N_t+1)$ null matrix and
$\left[\=I\right]$ is the  $(2N_t+1)\times(2N_t+1)$ identity matrix, the three non-null submatrixes on the right side
of Eq.~(\ref{Pmat}) are as follows:
\begin{eqnarray}
\left[\=P_{14}(z)\right]&=&\ko\left[\=I\right]-\frac{1}{\ko}\left[\=K_x\right]\cdot
\left[\=\epsilon(z)\right]^{-1}\cdot\left[\=K_x\right]\,,\\
\left[\=P_{32}(z)\right]&=&\frac{1}{\ko}\left[\=K_x\right]^2-\ko\left[\=\epsilon(z)\right]\,,\\
\left[\=P_{41}(z)\right]&=&\ko\left[\=\epsilon(z)\right]\,.
\end{eqnarray}
 
\subsection{Solution algorithm}
The column vectors $\left[\#f(0)\right]$ and $\left[\#f(d_3)\right]$ can be written using Eqs.~(\ref{einc})--(\ref{htr}) as  
\begin{equation}
\left[\#f(0)\right]=\left[
\begin{array}{cc}
\left[\=Y_e^+\right] & \left[\=Y_e^-\right]\\
\left[\=Y_h^+\right] & \left[\=Y_h^-\right]
 \end{array}
 \right]\cdot \left[
 \begin{array}{c}
 \left[\#A\right]\\
\left[\#R\right]
 \end{array}\right]\,,\qquad  
 \left[\#f(d_3)\right]=\left[
\begin{array}{c}
\left[\=Y_e^+\right] \\
\left[\=Y_h^+\right] 
 \end{array}
 \right]\cdot \left[\#T\right]\,, \label{bc0}
 \end{equation}
where
\begin{eqnarray}
\left[\#A\right]&=&\big[a_s^{(-N_t)},~a_s^{(-N_t+1)},~...,~a_s^{(0)},~...,~a_s^{(N_t-1)},~a_s^{(N_t)},\nonumber\\
&&~a_p^{(-N_t)}, ~a_p^{(-N_t+1)},~...,~a_p^{(0)},~...,~a_p^{(N_t-1)},~a_p^{(N_t)}\big]^T\,,\\
\left[\#R\right]&=&\big[r_s^{(-N_t)},~r_s^{(-N_t+1)},~...,~r_s^{(0)},~...,~r_s^{(N_t-1)},~r_s^{(N_t)},\nonumber\\
&&~r_p^{(-N_t)}, ~r_p^{(-N_t+1)},~...,~r_p^{(0)},~...,~r_p^{(N_t-1)},~r_p^{(N_t)}\big]^T\,,\\
\left[\#T\right]&=&\big[t_s^{(-N_t)},~t_s^{(-N_t+1)},~...,~t_s^{(0)},~...,~t_s^{(N_t-1)},~t_s^{(N_t)},\nonumber\\
&&~t_p^{(-N_t)}, ~t_p^{(-N_t+1)},~...,~t_p^{(0)},~...,~t_p^{(N_t-1)},~t_p^{(N_t)}\big]^T\,,
\end{eqnarray}
and the non-zero entries of $(4N_t+2)\times (4N_t+2)$-matrixes $\left[\=Y_{e,h}^\pm\right]$ are as follows:
\begin{eqnarray}
\left(Y_e^\pm\right)_{nm}=&1\,,&\quad n=m+2N_t+1\,,\\
\left(Y_e^\pm\right)_{nm}=&\mp{k_z^{(n)}\over\ko}\,, &\quad n=m-2N_t-1\,,\\
\left(Y_h^\pm\right)_{nm}=&\mp{k_z^{(n)}\over\ko}\,,&\quad n=m\in [1,~2N_t+1]\,,\\
\left(Y_h^\pm\right)_{nm}=&-1\,,&\quad n=m\in\left[2N_t+2,~4N_t+2\right]\,.
\end{eqnarray}

 
In order to devise a stable algorithm~\cite{Li93, Chateau94,Moharam95,Wang04},  the region $0\leq z\leq d_1$ is
divided into $N_d$ slices and  the region $d_1<z<d_2$ into $N_g$ slices, but the region $d_2\leq z\leq d_3$ is
kept as just one slice.  So, there are $N_d+N_g+1$ slices and $N_d+N_g+2$ interfaces. In the $j$th slice, $j\in\left[1,N_s+N_g+1\right]$, bounded by the
planes $z=z_{j-1}$ and $z=z_j$, we approximate
\begin{equation}
\left[\=P(z)\right] = \left[\=P\right]_j = \left[\=P\left(z_j+z_{j-1}\over 2\right)\right]\,,\quad
z\in\left(z_j,z_{j-1}\right)\,,
\end{equation}
so that  Eq.~(\ref{mode}) yields
\begin{equation}
\left[\#f(z_{j-1})\right]=\left[\=G\right]_j\cdot\exp\left\{-i\Delta_j\left[\=D\right]_j\right\}\cdot\left[\=G\right]_j^{-1}\cdot\left[\#f(z_j)\right]\,, 
\label{eqn41}
\end{equation}
where $\Delta_j=z_j-z_{j-1}$, $\left[\=G\right]_j$ is a square matrix comprising the eigenvectors of $\left[\=P\right]_j$ as its columns, and the diagonal matrix $\left[\=D\right]_j$ contains the eigenvalues of $\left[\=P\right]_j$ in the same order.

Let us define auxiliary column vectors $\left[\#T\right]_j$ and transmission matrixes $\left[\=Z\right]_j$ by the relation
\cite{Wang04}
\begin{equation}
\left[\#f(d_j)\right]=\left[\=Z\right]_j\cdot\left[\#T\right]_j\,,\qquad j\in\left[0,~N_d+N_g+1\right]\,,\label{eq50}
\end{equation}
where $d_0=0$,
\begin{equation}
\left[\#T\right]_{N_d+N_g+1}=\left[\#T\right]\,,\qquad \left[\=Z\right]_{N_d+N_g+1}=\left[
\begin{array}{c}
\left[\=Y_e^+\right]\\
\left[\=Y_h^+\right]
\end{array}
\right] \,.
\end{equation}
To find $\left[\#T\right]_j$ and $\left[\=Z\right]_j$ for $j\in\left[0,~N_d+N_g\right]$, we substitute Eq.~(\ref{eq50}) in (\ref{eqn41}), which results in the relation
\begin{eqnarray}
\left[\=Z\right]_{j-1}\cdot\left[\#T\right]_{j-1}=&&\left[\=G\right]_j\cdot\left[
\begin{array}{cc}
e^{-i\Delta_j[\=D]_j^u}&0\\
0&e^{-i\Delta_j[\=D]_j^l}
\end{array}\right]\cdot
\left[\=G\right]_j^{-1}\cdot\left[\=Z\right]_j\cdot\left[\#T\right]_j\,,\nonumber\\
&& j\in\left[1,~N_d+N_g+1\right]\,,\label{ZT}
\end{eqnarray}
where $\left[\=D\right]_j^u$ and $\left[\=D\right]_j^l$ are the upper and lower diagonal submatrixes of $\left[\=D\right]_j$, respectively, when the eigenvalues are arranged in decreasing order of the imaginary part. 

Since $\left[\#T\right]_j$ and $\left[\=Z\right]_j$ cannot be determined simultaneously from Eq.~(\ref{ZT}), let us define \cite{Wang04}
\begin{equation}
\left[\#T\right]_{j-1}=\exp\left\{-i\Delta_j\left[\=D\right]_j^u\right\}\cdot
\left[\=W\right]_j^u\cdot\left[\#T\right]_j\,,\label{TJ}
\end{equation}
where the square matrix $\left[\=W\right]_j^u$ and its counterpart $\left[\=W\right]_j^l$ are defined via
\begin{equation}
\left[
\begin{array}{c}
\left[\=W\right]_j^u\\
\left[\=W\right]_j^l
\end{array}
\right]=\left[\=G\right]_j^{-1}\cdot\left[\=Z\right]_j\,.\label{WU}
\end{equation}
Substitution of Eq.~(\ref{TJ}) in (\ref{ZT})  results in the relation
\begin{eqnarray}
\nonumber
&&\left[\=Z\right]_{j-1}=\left[\=G\right]_j\cdot\left[
\begin{array}{c}
\left[\=I\right]\\
\exp\left\{{-i\Delta_j[\=D]_j^l}\right\}\cdot
\left[\=W\right]_j^l\cdot\left\{\left[\=W\right]_j^u\right\}^{-1}\cdot
\exp\left\{{i\Delta_j[\=D]_j^u}\right\}
\end{array}
\right]\,,
\\[5pt]
&&
\qquad \qquad{j}\in\left[1,N_d+N_g+1\right]\,.\label{Z}
\end{eqnarray}
From Eqs.~(\ref{WU}) and (\ref{Z}), we find  $\left[\=Z\right]_0$ in terms of $\left[\=Z\right]_{N_d+N_g+1}$. After partitioning 
\begin{equation}
\left[\=Z\right]_0=\left[
\begin{array}{c}
\left[\=Z\right]_0^u\\
\left[\=Z\right]_0^l
\end{array}\right]\,,
\end{equation}
and using Eqs.~(\ref{bc0}) and (\ref{eq50}), $\left[\#R\right]$ and $\left[\#T\right]_0$ are found as follows:
\begin{equation}
\left[
\begin{array}{c}
\left[\#T\right]_0\\
\left[\#R\right]
\end{array}
\right]=\left[
\begin{array}{ccc}
\left[\=Z\right]_0^u & &-\left[\=Y_e^-\right]\\
\left[\=Z\right]_0^l & &-\left[\=Y_h^-\right]
\end{array}
\right]^{-1}\cdot\left[
\begin{array}{c}
\left[\=Y_e^+\right]\\
\left[\=Y_h^+\right]
\end{array}
\right]\cdot
\left[\#A\right]\,.\label{T0R}
\end{equation}

Equation~(\ref{T0R}) is obtained by enforcing the usual boundary conditions across the plane $z=0$. After $\left[\#T\right]_0$ is known, $\left[\#T\right]=\left[\#T\right]_{N_d+N_g+1}$ is found by reversing the sense of iterations in Eq.~(\ref{TJ}).

\section{Numerical Results and Discussion}\label{nrd}

\subsection{Homogeneous dielectric partnering material}
Let us begin with the dielectric partnering material being homogeneous, i.e., $\epsilon_d(z)$ is independent of $z$. This case has been numerically illustrated by Homola~\cite[p.~38]{Homola_book} and we adopted the same parameters: ${\lambdao}=800$~nm, $\eps_d=1.766$ (water), $\eps_m=-25+1.44i$ (gold),  and $L=672$~nm. The incident plane wave is $p$
polarized ($a_p\pn= \delta_{n0}$ and $a_s\pn\equiv 0\,\forall{n\in{\mathbb Z}}$) and the quantity of importance is the absorbance
\begin{equation}
A_p=1-\sum_{n=-N_t}^{N_t}\left(\left|r_s^{(n)}\right|^2+\left|r_p^{(n)}\right|^2+\left|t_s^{(n)}\right|^2+\left|t_p^{(n)}\right|^2\right){\rm Re}\left[k_z\pn/k_z^{(0)}\right]\,,
\end{equation}
which simplifies to
\begin{equation}
A_p=1-\sum_{n=-N_t}^{N_t}\left( \left|r_p^{(n)}\right|^2 +\left|t_p^{(n)}\right|^2\right){\rm Re}\left[k_z\pn/k_z^{(0)}\right]\,,
\end{equation}
because all materials are isotropic.

\begin{figure}[!ht]
\begin{center}
\includegraphics[width=6in]{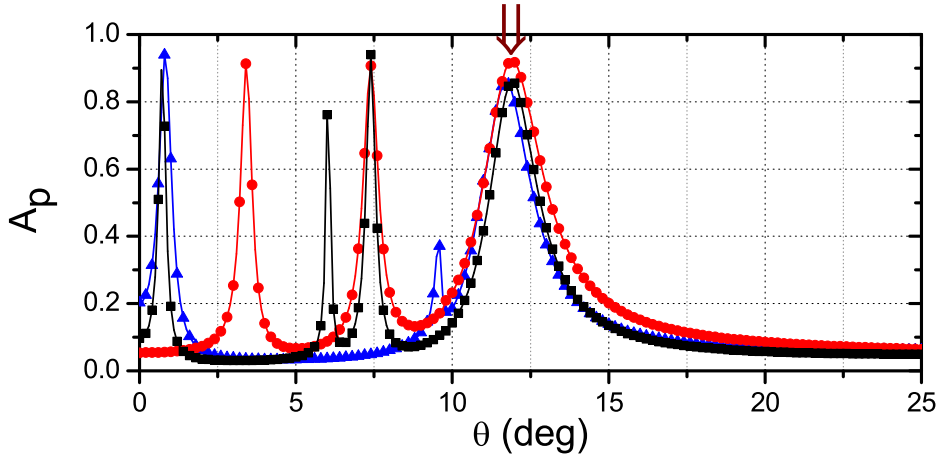}
\caption{Absorbance $A_p$ as a function of the incidence angle $\theta$ for a   sinusoidal surface-relief grating defined by Eq.~(\ref{sinusoidal}).  Black squares represent $d_1=1500$~nm, red circles $d_1=1000$~nm, and blue triangles $d_1=800$~nm. The grating depth ($d_2-d_1=50$~nm) and the thickness of the metallic layer ($d_3-d_2=30$~nm) are the same for all three cases. The arrow identifies
 an SPP wave.}
\label{sinusoidalfig}
\end{center}
\end{figure}

\begin{figure}[!ht]
\begin{center}
\includegraphics[width=6in]{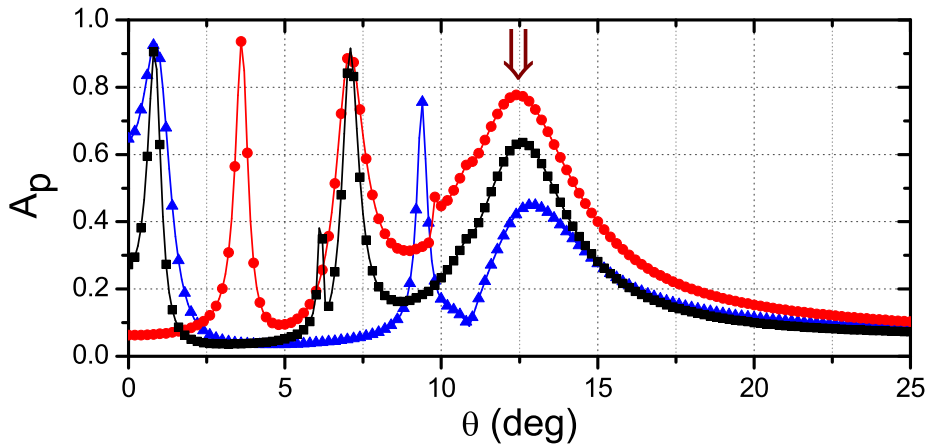}
\caption{Same as Fig.~\ref{sinusoidalfig} except that the surface-relief grating is defined by Eq.~(\ref{ourgrating}) with $L_1=0.5L$.}
\label{ours}
\end{center}
\end{figure}

Figure~\ref{sinusoidalfig} shows the variation of $A_p$ versus the incidence angle $\theta$ for a sinusoidal surface-relief grating    defined by
\cite{Homola_book}
 \begin{equation}
g(x)=\frac{1}{2}(d_2-d_1)\left[1+\sin \left(\frac{2\pi x}{L}\right)\right]
\label{sinusoidal}
\end{equation} 
and Fig.~\ref{ours} shows the  same for the surface-relief grating   defined by Eq.~(\ref{ourgrating}) with $L_1=0.5L$. For computational purposes, we set $N_d=1$,  $N_g=50$, and $N_t=10$, after ascertaining that the reflectances $\left|r_p^{(n)}\right|^2{\rm Re}\left[k_z\pn/k_z^{(0)}\right]$ and the transmittances $\left|t_p^{(n)}\right|^2{\rm Re}\left[k_z\pn/k_z^{(0)}\right]$ converged for all $n\in\left[-N_t,N_t\right]$.
 
Each figure shows plots of $A_p$ vs. $\theta$ for three different values of the thickness $d_1$, in order to distinguish~\cite{PartI} between
\begin{itemize}
\item[(i)] surface waves \cite{Faryadjosab}, which must be independent of $d_1$ for sufficiently large values of that parameter,
and
\item[(ii)] waveguide modes \cite{KBbook,Mbook}, which must depend on $d_1$
\end{itemize}
as has been shown elsewhere \cite{Otto1971,Faryadpra}. In both figures, an absorbance peak at $\theta\simeq 12^\circ$ for all three values of $d_1$ indicates the excitation of an SPP wave. 

The relative wavenumbers $k_x\pn/\ko$ of a few Floquet harmonics   at $\theta=12^\circ$ are given in Table~\ref{table1}. The solution of the canonical boundary-value problem (when both partnering materials are semi-infinite along the $z$ axis and their interface
is planar) \cite{Homola_book} shows that the relative wavenumber $\kappa/\ko$ of the SPP wave that can be guided by the planar gold-water interface is 
\begin{equation}
\kappa/\ko=\sqrt{\eps_d\eps_m/(\eps_d+\eps_m)}=1.3784+0.0030i\,.\label{goldkappa}
\end{equation}
A comparison of Table~\ref{table1} and Eq.~(\ref{goldkappa}) confirms that an SPP wave is excited at $\theta=12^\circ$ as the Floquet harmonic of order $n=+1$. We note that the absorbance peak in Fig.~\ref{ours} is not only wider than in Fig.~\ref{sinusoidalfig}, but also of lower magnitude, which points out the critical importance of the shape function $g(x)$ of the
surface-relief grating. The incidence angle determined by Homola~\cite[p.~38]{Homola_book} is approximately $11^\circ$, the
small difference  between his and our results being  (i)
due to the different methods of computation and (ii)
the fact that, while Homola had semi-infinite dielectric and metallic partnering materials, we have the two of finite thickness.

 \begin{table}[!ht]
\begin{center}
\caption{Relative wavenumbers $k_x\pn/\ko$ of Floquet harmonics for a gold-water grating when $\theta=12^\circ$.
Boldface entries signify SPP waves.}
\begin{tabular}{||c|c|c|c|c|c||}\hline\hline
  & $n=-2$ &  $n=-1$  & $n=0$ & $n=1$  & $n=2$ \\\hline
$\theta=12^\circ$  & $-2.1730$ &  $-0.9826$  & $0.2079$ & $\mathbf{ 1.3984}$ & $2.5889$ \\\hline\hline
\end{tabular}\label{table1}
\end{center}
\end{table}

\subsection{Periodically nonhomogeneous dielectric partnering material}\label{3.B}

Now let us move on to the   excitation of {\it multiple} SPP waves by  a surface-relief grating where the dielectric partnering material has a periodic nonhomogeneity normal to the mean plane of the surface-relief grating:  
\begin{equation}
\epsilon_d(z)=\left[\left({n_b+n_a\over 2}\right)+\left({{n_b-n_a}\over 2}\right)\sin\left(\pi{d_2-z\over \Omega}\right)\right]^2\,,\quad z> 0\,,
\label{rugate1}
\end{equation}
where 2$\Omega$ is the period. We chose $n_a=1.45$ and $n_b=2.32$ from an example provided by Baumeister~\cite[Sec. 5.3.3.2 ]{Baumeister}. 
For all calculations reported in the remainder of this paper, we chose the metal to be  bulk aluminum ($\epsmet=-56+i21$) and the free-space wavelength ${\lambdao}=633$~nm. The surface-relief grating is defined by Eq.~(\ref{ourgrating}) with $L_1=0.5L$. We fixed $N_t=8$ after ascertaining that the absorbances for $N_t=8$ converged to within $<1~\%$ of the absorbances calculated with $N_t=9$. The grating depth $d_2-d_1=50$~nm and the  thickness $d_3-d_2=30$~nm were also fixed, as their variations would not qualitatively affect the excitation of multiple SPP waves. Numerical results for $\Omega=\lambdao$ and $\Omega=1.5\lambdao$ are now presented.

\subsubsection{$\Omega=\lambdao$}\label{3.B.1}

\begin{figure}[!ht]
\begin{center}
\includegraphics[width=5in]{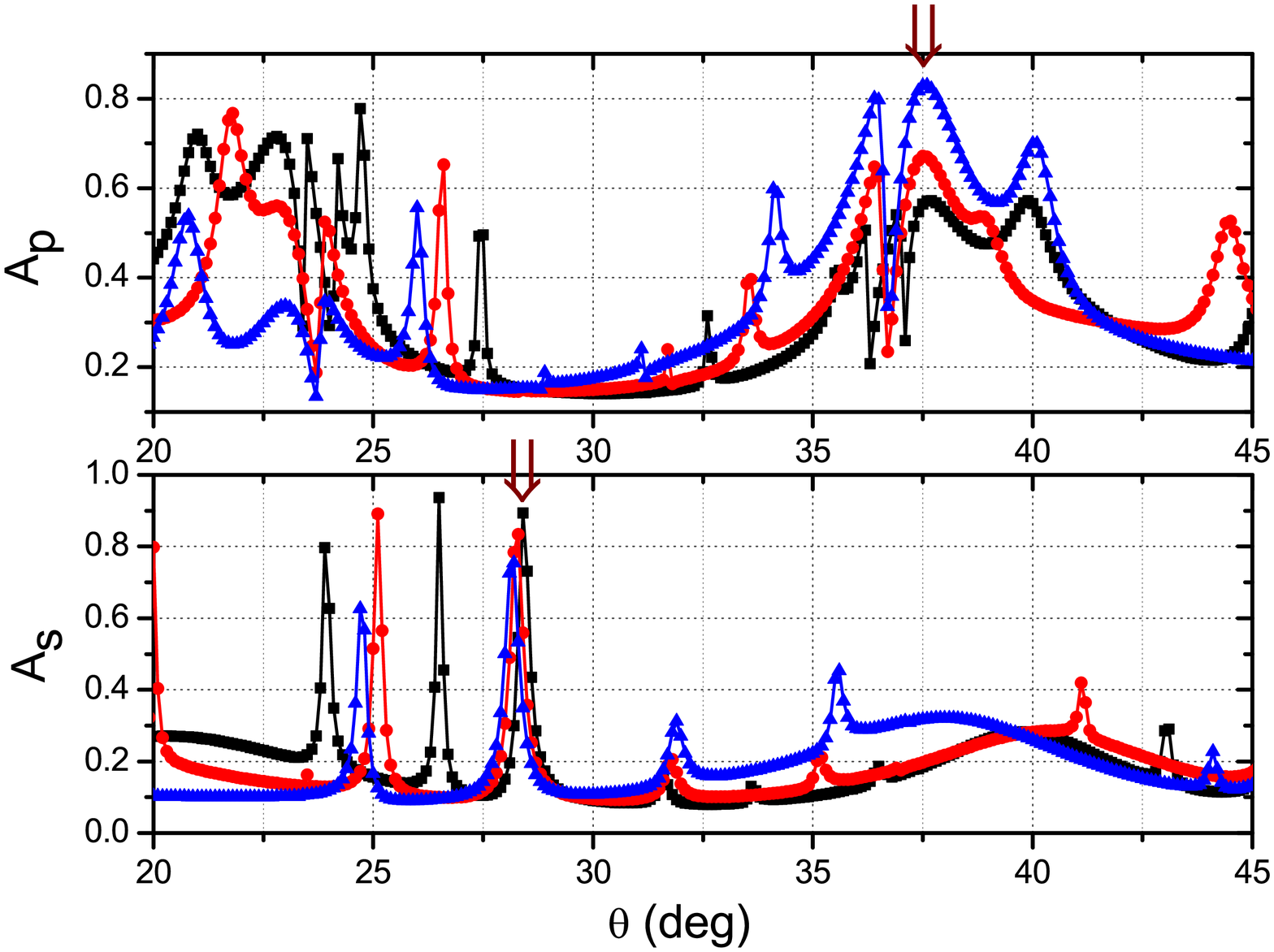}
\caption{Absorbances $A_p$ and $A_s$ as functions of the incidence angle $\theta$ for a surface-relief grating defined by Eq.~(\ref{ourgrating}) with $L_1=0.5L$, when  ${\lambdao}= 633$~nm, $\Omega={\lambdao}$, and $L=\lambdao$. Black squares are for $d_1=6\lambdao$, red circles for $d_1=5\lambdao$, and blue triangles for $d_1=4\lambdao$.  The grating depth ($d_2-d_1=50$~nm) and the thickness of the metallic layer ($d_3-d_2=30$~nm) are the same for all plots. Each arrow identifies
 an SPP wave.  }
\label{fig1L1}
\end{center}
\end{figure}

\begin{figure}[!ht]
\begin{center}
\includegraphics[width=5in]{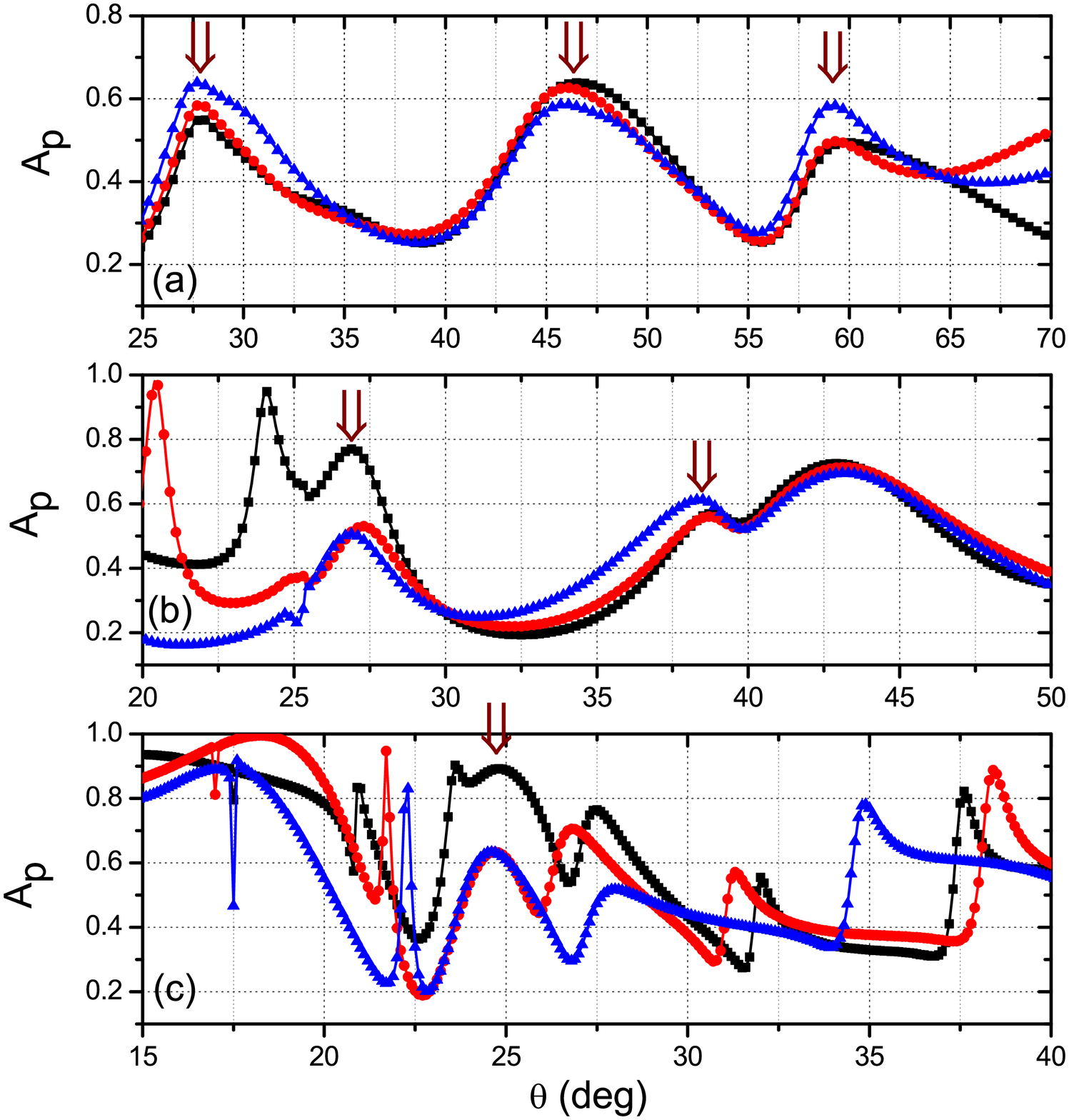}
\caption{Absorbance $A_p$ as a function of the incidence angle $\theta$ for a surface-relief grating defined by Eq.~(\ref{ourgrating}) with $L_1=0.5L$, when  ${\lambdao}= 633$~nm, $\Omega={\lambdao}$; and (a) $L=0.75\lambdao$, (b) $L=0.7\lambdao$, and (c) $L=0.55\lambdao$. Black squares are for $d_1=6\lambdao$, red circles for $d_1=5\lambdao$, and blue triangles for $d_1=4\lambdao$.  The grating depth ($d_2-d_1=50$~nm) and the thickness of the metallic layer ($d_3-d_2=30$~nm) are the same for all plots. Each arrow identifies
 an SPP wave.}
\label{fig1Lap}
\end{center}
\end{figure}

Let us commence with $\Omega=\lambdao$. The solution of the corresponding canonical boundary-value problem results in five $p$-polarized and one $s$-polarized SPP waves~\cite{Faryadjosab}, the relative wavenmbers $\kappa/\ko$ being provided in Table~\ref{table2}. To analyze the excitation of $s$-polarized SPP waves in the grating-coupled configuration, we calculated the absorbance   
\begin{equation}
A_s=1-\sum_{n=-N_t}^{n=N_t}\left(\left|r_s^{(n)}\right|^2+ \left|t_s^{(n)}\right|^2\right){\rm Re}[k_z\pn/k_z^{(0)}]\,
\end{equation}
for $a_s\pn= \delta_{n0}$ and $a_p\pn\equiv 0\,\forall{n\in{\mathbb Z}}$.
Both $A_p$ and $A_s$ were calculated as functions of $\theta$ for  $d_1\in\left\{4\lambdao,5\lambdao,6\lambdao\right\}$, with
$N_g$ and $N_d$ selected to have slices of thickness  $2$~nm in the region
$0\leq{z}\leq{d_1}$ but $1$~nm in the region $d_1<z<d_2$.
Figures~\ref{fig1L1}, \ref{fig1Lap}(a), \ref{fig1Lap}(b), and \ref{fig1Lap}(c) present the absorbances as functions of $\theta$ for $L=\lambdao$, $0.75\lambdao$, $0.7\lambdao$, and $0.55\lambdao$, respectively.

 \begin{table}[!ht]
\begin{center}
\caption{Relative wavenumbers $\kappa/\ko$ of possible SPP waves obtained by the solution of the 
canonical boundary-value problem \cite{Faryadjosab} for  $\Omega=\lambdao$. Other parameters are provided in the beginning of Sec.~\ref{3.B}.}
\begin{tabular}{||c||c|c|c|c|c||}\hline\hline
{\bf $s$-pol}  & $1.48639+0.00132i$ &  & & & \\\hline
 {\bf $p$-pol} & $1.36479+0.00169i$ & $1.61782+0.00548i$ & $1.87437+0.00998i$ &  $2.06995+0.01526i$ &  $2.21456+0.00246i$ \\\hline\hline
\end{tabular}\label{table2}
\end{center}
\end{table}

 \begin{table}[!ht]
\begin{center}
\caption{Relative wavenumbers $k_x\pn/\ko$ of Floquet harmonics when $L=\lambdao$. Boldface entries signify SPP waves.}
\begin{tabular}{||c|c|c|c|c|c||}\hline\hline
  & $n=-2$ &  $n=-1$  & $n=0$ & $n=1$  & $n=2$ \\\hline
 $\theta=28^\circ$ & $-1.5305$ &  $-0.5305$  & $0.4695$ & $\mathbf{1.4695}$ & $2.4695$ \\\hline
 $\theta=37.5^\circ$ & $ -1.3912$ &  $-0.3912$  & $0.6088$ & $\mathbf{1.6088}$ & $2.6088$ \\\hline\hline
\end{tabular}\label{table3}
\end{center}
\end{table}

For all three
values of $d_1$,  a peak is present at $\theta=37.5^\circ$ in
the plots of $A_p$ vs. $\theta$ in Fig.~\ref{fig1L1}. The relative wavenumbers $k_x\pn/\ko$ of several Floquet harmonics at this incidence angle  are given in Table~\ref{table3}. At $\theta=37.5^\circ$,  $k_x^{(1)}/\ko=1.6088$ is close to ${\rm Re}[\kappa/\ko]=1.61782$, where $\kappa/\ko$ is the relative wavenumber of an SPP wave in the canonical boundary-value problem \cite{Faryadjosab} as provided in Table~\ref{table2}; likewise, $k_x^{(-2)}/\ko=1.3912$  is close to ${\rm Re}[\kappa/\ko]=1.36479$
(Table~\ref{table2}) for another solution of the canonical boundary-value problem. Thus, 
the $A_p$-peak could represent the grating-coupled excitation of either one or two $p$-polarized SPP waves. To resolve this issue, we computed $A_p$ vs. $\theta$ for $L=1.02\lambdao$ (not shown). For the new grating period, the $A_p$-peak shifted to a higher value of $\theta$. This shift showed that the SPP wave at the $A_p$-peak in Fig.~\ref{fig1L1} is excited as a Floquet harmonic with a positive index $n$ and not the one with a negative index. So the $A_p$-peak represents the excitation of a $p$-polarized SPP wave by a Floquet harmonic of order $n=1$ with $k_x^{(1)}/\ko=1.6088$.

A peak is also present at $\theta=28^\circ$ in
the plots of $A_s$ vs. $\theta$, for all three
values of $d_1$ in Fig.~\ref{fig1L1}. The solution of the canonical boundary-value problem
\cite{Faryadjosab} indicates the excitation of an $s$-polarized SPP wave when $\kappa/\ko=1.48639+0.00132$.
The real part of this relative wavenumber is close to $k_x^{(1)}/\ko=1.4695$ for $\theta=28^\circ$; see Table~\ref{table3}. So the $A_s$-peak represents the grating-coupled excitation of an $s$-polarized SPP wave.

 \begin{table}[!ht]
\begin{center}
\caption{Same as Table~\ref{table3} except for  $L=0.75\lambdao$.}
\begin{tabular}{||c|c|c|c|c|c||}\hline\hline
  & $n=-2$ &  $n=-1$  & $n=0$ & $n=1$  & $n=2$\\\hline
 $\theta=32.5^\circ$ & $-2.1294$ &  $-0.7960$  & $0.5373$ & $\mathbf{ 1.8760}$ & $3.2040$ \\\hline
 $\theta=51^\circ$ & $\mathbf{-1.8895}$ &  $-0.5561$  & $0.7771$ & ${ 2.1104}$ & $3.4438$ \\\hline
 $\theta=64^\circ$ & $-1.7679$ &  $-0.4345$  & $0.8988$ & $\mathbf{ 2.2321}$ & $3.5655$ \\\hline\hline
\end{tabular}\label{table4}
\end{center}
\end{table}

Since not all possible $p$-polarized SPP waves (predicted after the solution of the canonical boundary-value problem) can  be excited with period $L=\lambdao$ of the surface-relief grating, the grating period needs to be changed to excite the remaining SPP waves. The plots of $A_p$ vs. $\theta$ for $L=0.75\lambdao$ are presented in Fig.~\ref{fig1Lap}(a), again  for  $d_1\in\left\{4\lambdao,5\lambdao,6\lambdao\right\}$. The figure shows three $A_p$-peaks at $\theta=32.5^\circ$, $51^\circ$, and $64^\circ$ that are present for all three chosen values of $d_1$. The relative wavenumbers of several Floquet harmonics at these values of the incidence angle are given in Table~\ref{table4}. The $A_p$-peak at $\theta=32.5^\circ$ represents the excitation of a $p$-polarized SPP wave as a   Floquet harmonic  of order $n=1$ because $k_x^{(1)}/\ko=1.8760$ is close to  ${\rm Re}\left[1.87437+0.00998i\right]$ in Table~\ref{table2}. Remarkably, the $A_p$-peak at $\theta=51^\circ$ represents the excitation of the same SPP wave as a Floquet harmonic  of order $n=-2$. Finally, the $A_p$-peak at $\theta=64^\circ$ represents the excitation of another $p$-polarized SPP wave as a Floquet harmonic of order $n=1$.

 \begin{table}[!ht]
\begin{center}
\caption{Same as Table~\ref{table3} except for  $L=0.7\lambdao$.}
\begin{tabular}{||c|c|c|c|c|c||}\hline\hline
  & $n=-2$ &  $n=-1$  & $n=0$ & $n=1$  & $n=2$\\\hline
 $\theta=27^\circ$ & $-2.4031$ &  $-0.9746$  & $0.4534$ & $\mathbf{ 1.8826}$ & $3.3111$ \\\hline
 $\theta=38^\circ$ & $-2.2415$ &  $-0.8129$  & $0.6157$ & $\mathbf{ 2.0442}$ & $3.4728$ \\\hline\hline
\end{tabular}\label{table5}
\end{center}
\end{table}

Two $A_p$-peaks, at $\theta=27^\circ$ and $38^\circ$, are present for all values of $d_1$ in Fig.~\ref{fig1Lap}(b) for $L=0.7\lambdao$. The relative wavenumbers of several Floquet harmonics at these two values of the incidence angle are given in Table~\ref{table5}. The $A_p$-peak at $\theta=27^\circ$ represents the excitation of a $p$-polarized SPP wave because $k_x^{(1)}/\ko=1.8826$ at this incidence angle is close to ${\rm Re}[\kappa/\ko]=1.87437$ (Table~\ref{table2}) of a $p$-polarized SPP wave predicted by the canonical problem.  The $A_p$-peak at $\theta=38^\circ$ represents another $p$-polarized SPP wave because $k_x^{(1)}/\ko=2.0442$ is close to  ${\rm Re}[\kappa/\ko]=2.06995$ (Table~\ref{table2}).

 \begin{table}[!ht]
\begin{center}
\caption{Same as Table~\ref{table3} except for  $L=0.55\lambdao$.}
\begin{tabular}{||c|c|c|c|c|c||}\hline\hline
  & $n=-2$ &  $n=-1$  & $n=0$ & $n=1$  & $n=2$\\\hline
 $\theta=25^\circ$ & $-3.2137$ &  $\mathbf{ -1.3956}$  & $0.4226$ & $ 2.2408$ & $4.0590$ \\\hline\hline
\end{tabular}\label{table6}
\end{center}
\end{table}

All the SPP waves that can be guided by the planar interface of the rugate filter (for the chosen value of $\Omega$) and aluminum have been shown to be excited by the grating-coupled configuration except the one with $\kappa/\ko=1.36479+0.00169i$ (Table~\ref{table2}). This $p$-polarized SPP wave was found to be excited with a
surface-relief grating of period $L=0.55\lambdao$. The variation of $A_p$ vs. $\theta$ for this case is shown in Fig.~\ref{fig1Lap}(c). The $A_p$-peak at $\theta=25^\circ$ is independent of the thickness $d_1$ of the rugate filter. The relative wavenumbers $k_x\pn/\ko$ of several Floquet harmonics at this value of the incidence angle are given in Table~\ref{table6}. Comparing Tables~\ref{table2} and \ref{table6}, we see that $k_x^{(-1)}/\ko$ is close to ${\rm Re}\left[1.36479+0.00169i\right]$ and $k_x^{(1)}/\ko$ is close to ${\rm Re}\left[2.21456+0.00246i\right]$. This suggests that either of the two or both SPP waves are excited at $\theta=25^\circ$. To resolve this issue, we computed $A_p$ vs. $\theta$ for $L=0.57\lambdao$ (not shown) and the $A_p$-peak shifted to a smaller value of the incidence angle. This shift indicated that the absorbance peak at $\theta=25^\circ$ represents an SPP wave due to a Floquet harmonic of negative order. So the $A_p$-peak at $\theta=25^\circ$ represents the excitation of the SPP wave as a Floquet harmonic of order $n=-1$.

\subsubsection{$\Omega=1.5\lambdao$}\label{3.B.2}

The relative wavenumbers of possible SPP waves that can be guided by the {\it planar} interface of the chosen rugate filter and the metal are given in Table~\ref{table7} for $\Omega=1.5\lambdao$. In this case, the solution of the canonical
boundary-value problem indicated that multiple $s$-polarized SPP waves can also be guided in addition to multiple $p$-polarized SPP waves. 

\begin{figure}[!ht]
\begin{center}
\includegraphics[width=5in]{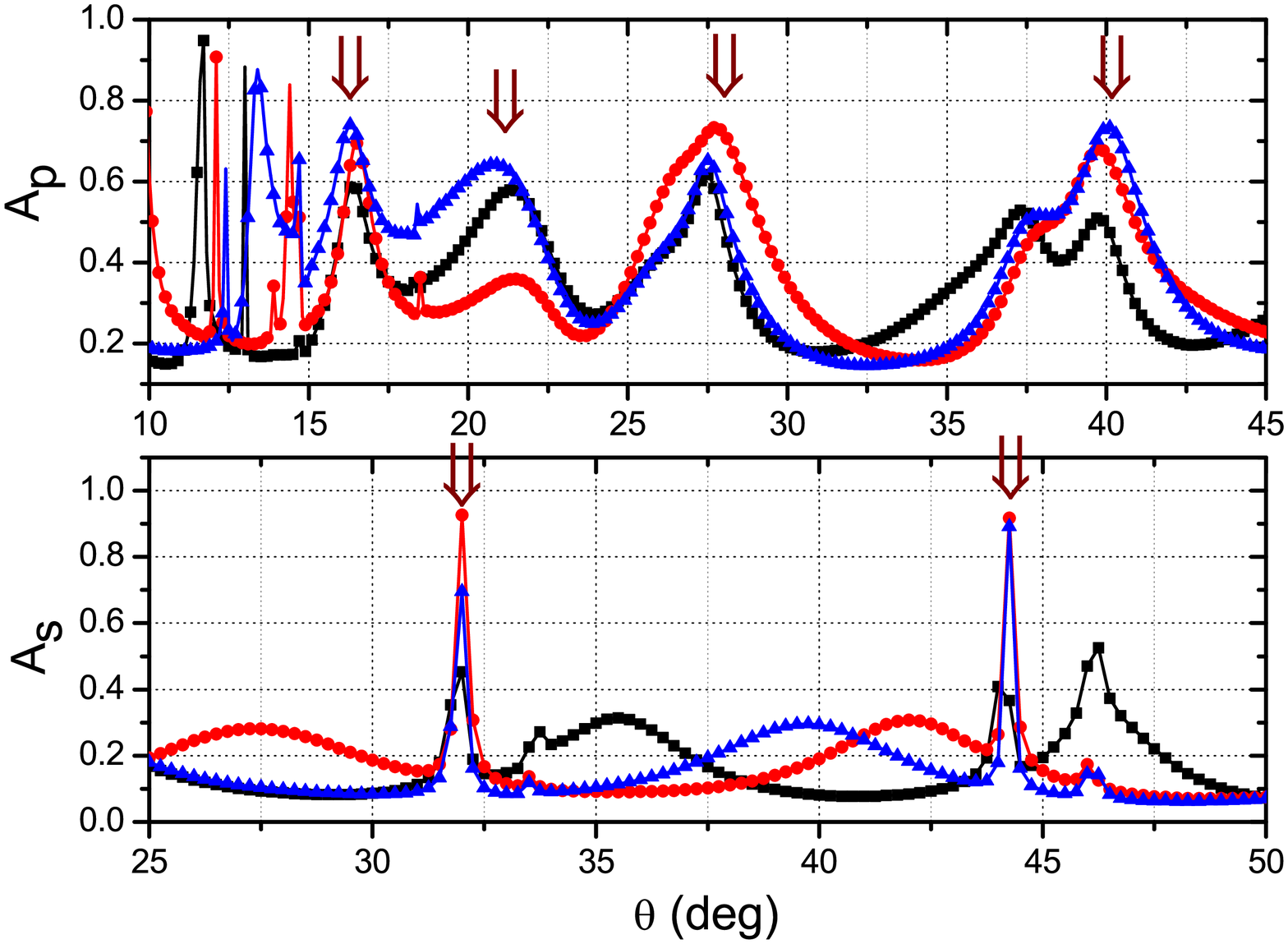}
\caption{Absorbances $A_p$ and $A_s$ as functions of the incidence angle $\theta$ for a surface-relief grating defined by Eq.~(\ref{ourgrating}) with $L_1=0.5L$,  when ${\lambdao}= 633$~nm, $\Omega=1.5{\lambdao}$, and $L=0.8\lambdao$. Black squares are for $d_1=9\lambdao$, red circles for $d_1=7.5\lambdao$, and blue triangles for $d_1=6\lambdao$. The grating depth ($d_2-d_1=50$~nm) and the width of the metallic layer ($d_3-d_2=30$~nm) are the same for all the plots. Each arrow indicates an SPP wave.}
\label{fig15L8}
\end{center}
\end{figure}

\begin{figure}[!ht]
\begin{center}
\includegraphics[width=5in]{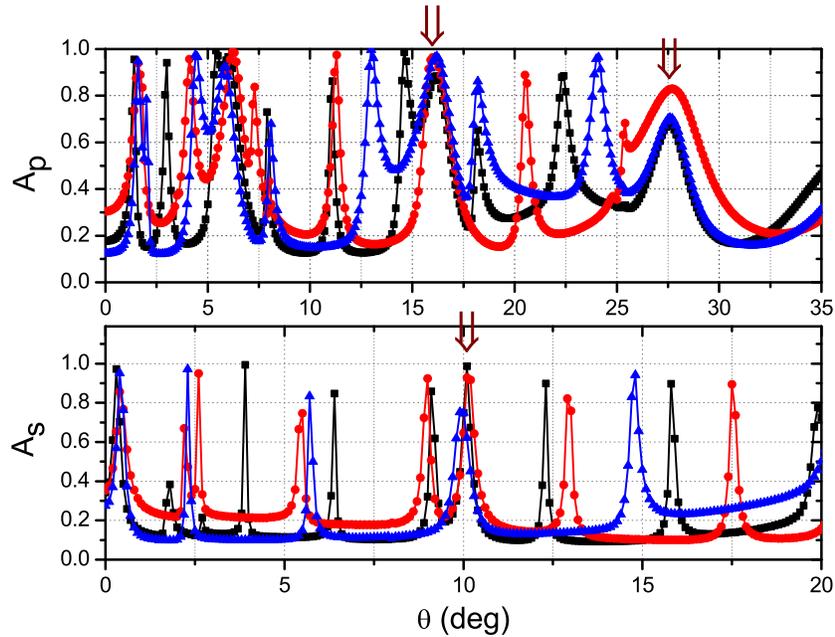}
\caption{Same as Fig.~\ref{fig15L8} except for $L=0.7\lambdao$.}
\label{fig15L7}
\end{center}
\end{figure}

\begin{figure}[!ht]
\begin{center}
\includegraphics[width=5in]{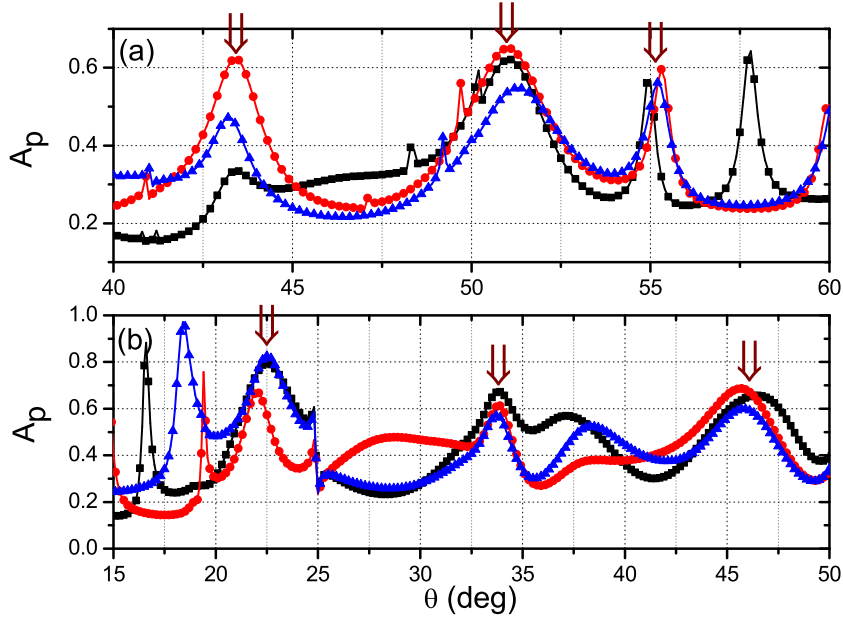}
\caption{Absorbance $A_p$ as a function of the incidence angle $\theta$ for a surface-relief grating defined by Eq.~(\ref{ourgrating}) with $L_1=0.5L$,  when ${\lambdao}= 633$~nm, $\Omega=1.5{\lambdao}$; and (a) $L=0.9\lambdao$ and (b) $L=0.55\lambdao$. Black squares are for $d_1=9\lambdao$, red circles for $d_1=7.5\lambdao$, and blue triangles for $d_1=6\lambdao$. The grating depth ($d_2-d_1=50$~nm) and the width of the metallic layer ($d_3-d_2=30$~nm) are the same for all the plots. Each arrow indicates an SPP wave.}
\label{fig15Lap}
\end{center}
\end{figure}

Absorbances $A_p$ and $A_s$, calculated for $d_1\in\left\{6\lambdao,7.5\lambdao,9\lambdao\right\}$ and $\theta\in\left[0^\circ,90^\circ\right)$, are presented in Figs.~\ref{fig15L8}, \ref{fig15L7}, \ref{fig15Lap}(a), and \ref{fig15Lap}(b) for $L=0.8\lambdao$, $0.7\lambdao$, $0.9\lambdao$, and $0.75\lambdao$, respectively. For the computations, the region $d_1<z<d_2$ was again divided into $1$-nm-thick slices; however, the region $0\leq z\leq d_1$ was divided into $3$-nm-thick slices to reduce the computation time, after ascertaining that the accuracy of the computed reflectances and transmittances had not been adversely affected.

 \begin{table}[!ht]
\begin{center}
\caption{Same as Table~\ref{table2} except for $\Omega=1.5\lambdao$.}
\begin{tabular}{||c||c|c|c||}\hline\hline
{\bf $s$-pol}   & $1.61507+0.00114$ & $1.78735+0.00078i$  &   \\\hline\hline
 {\bf $p$-pol} & $1.40725+0.00052i$    & $1.54121+0.00374i$ & $1.71484+0.0049i$  
  \\\cline{2-4} 
& $1.88541+0.00739i$ & $2.11513+0.0045i$ & $2.02159+0.01301i$ \\\hline\hline
\end{tabular}\label{table7}
\end{center}
\end{table}

 \begin{table}[!ht]
\begin{center}
\caption{Same as Table~\ref{table3} except for $L=0.8\lambdao$.}
\begin{tabular}{||c|c|c|c|c|c||}\hline\hline
  & $n=-2$ &  $n=-1$  & $n=0$ & $n=1$  & $n=2$\\\hline
$\theta=17^\circ$ & $-2.20276$ &  $-0.9576$  & $0.2924$ & $\mathbf{ 1.5424}$ & $2.7924$ \\\hline
$\theta=21^\circ$ & $\mathbf{ -2.1416}$ &  $-0.8916$  & $0.3584$ &  $1.6084$ & $2.8584$ \\\hline
$\theta=27.5^\circ$ & ${ -2.0382}$ &  $-0.7882$  & $0.4617$ & $\mathbf{ 1.7118}$ & $2.9618$\\\hline
$\theta=32^\circ$ & $-1.9701$ &  $-0.7201$  & $0.5299$ & $\mathbf{ 1.7799}$ & $3.0299$ \\\hline
$\theta=40^\circ$ & ${ -1.8572}$ &  $-0.6072$  & $0.6428$ & $\mathbf{ 1.8928}$ & $3.1428$ \\\hline
 $\theta=44^\circ$ & $\mathbf{-1.8053}$ &  $-0.5553$  & $0.6747$ & ${ 1.9447}$ & $3.1947$ \\\hline\hline
\end{tabular}\label{table8}
\end{center}
\end{table}

In the plots of $A_p$ vs. $\theta$ in Fig.~\ref{fig15L8}, the excitation of $p$-polarized SPP waves is indicated  at four values of the incidence angle: $\theta=17^\circ$, $21^\circ$, $27.5^\circ$, and $40^\circ$. The relative wavenumbers $k_x\pn/\ko$ of a few Floquet harmonics at these values of the incidence angle are given in Table~\ref{table8}. The $A_p$-peak at $\theta=17^\circ$ represents the excitation of a $p$-polarized SPP wave, because $k_x^{(1)}/\ko=1.5424$ is close to ${\rm Re}\left[\kappa/\ko\right]=1.54121$ (Table~\ref{table7}), where $\kappa/\ko$ is a solution of the canonical boundary-value problem. The $A_p$-peak at $\theta=21^\circ$ also represents a $p$-polarized SPP wave because  $k_x^{(-2)}/\ko=2.1416$ is close to ${\rm Re}\left[\kappa/\ko\right]=2.11513$ (Table~\ref{table7}), which is another solution of the canonical problem. The relative wavenumber $k_x^{(1)}/\ko=1.6084$ at $\theta=21^\circ$ was ruled out, after examining the plots for $L=0.82\lambdao$ (not shown). Similarly, the peak at $\theta=27.5^\circ$ is due to the excitation of another $p$-polarized SPP wave as a Floquet harmonic of order $n=1$ and not of the order $n=-2$. At $\theta=40^\circ$, both $k_x^{(1)}/\ko=1.8928$ and $k_x^{(-2)}/\ko=1.8572$ are close to ${\rm Re}\left[1.88541+0.00739i\right]$ (Table~\ref{table7}). However, the plots
of $A_p$ vs. $\theta$ for $L=0.82\lambdao$ (not shown) made us conclude that the SPP wave is excited as the Floquet harmonic of order $n=1$.

In the plots of $A_s$ vs. $\theta$ in Fig.~\ref{fig15L8}, two peaks at $\theta=32^\circ$ and $44^\circ$ are present for all values of $d_1$. The $A_s$-peak at $\theta=32^\circ$ represents the excitation of an $s$-polarized SPP wave because $k_x^{(1)}/\ko=1.7799$ is close to ${\rm Re}\left[\kappa/\ko\right]=1.78735$ (Table~\ref{table7}), which is a solution of the canonical 
boundary-value problem for an $s$-polarized SPP wave. The other $A_s$-peak at $\theta=44^\circ$ represents the excitation of the same $s$-polarized SPP wave as a  Floquet harmonic of order $n=-2$.

 \begin{table}[!ht]
\begin{center}
\caption{Same as Table~\ref{table3} except for  $L=0.7\lambdao$.}
\begin{tabular}{||c|c|c|c|c|c||}\hline\hline
 & $n=-2$ &  $n=-1$  & $n=0$ & $n=1$  & $n=2$\\\hline
 $\theta=10^\circ$ & $-2.6835$ &  $-1.2549$  & $0.1736$ & $\mathbf{ 1.6022}$ & $3.0308$ \\\hline
 $\theta=16^\circ$ & $-2.5815$ &  $-1.1529$  & $0.2756$ & $\mathbf{ 1.7042}$ & $3.1328$ \\\hline
 $\theta=27.5^\circ$ & $-2.3954$ &  $-0.9668$  & $0.4617$ & $\mathbf{ 1.8903}$ & $3.3189$ \\\hline\hline
\end{tabular}\label{table9}
\end{center}
\end{table}

For $L=0.7\lambdao$, the absorbances $A_p$ and $A_s$ are presented as
functions of $\theta$   in Fig.~\ref{fig15L7}. The relative wavenumbers $k_x^{(n)}/\ko$ of several Floquet harmonics at those values of the incidence angles where peaks are present independent of the value of $d_1$ are given in Table~\ref{table9}.
In the plots of $A_p$ vs. $\theta$, the peak at $\theta=16^\circ$ represents the excitation of a $p$-polarized SPP wave because $k_x^{(1)}/\ko=1.7042$ is close to ${\rm Re}[\kappa/\ko]=1.71484$ (Table~\ref{table7}). Similarly, the $A_p$-peak at $\theta=27.5^\circ$ represents the excitation of a $p$-polarized SPP wave, as $k_x^{(1)}/\ko=1.8903$ is close to ${\rm Re}[\kappa/\ko]=1.88541$ (Table~\ref{table7}). In the plots of $A_s$ vs. $\theta$, the peak at $\theta=10^\circ$ represents the excitation of an $s$-polarized SPP wave because $k_x^{(1)}/\ko=1.6022$ at this angle is close to ${\rm Re}[\kappa/\ko]=1.61507$ (Table~\ref{table7}). 

 \begin{table}[!ht]
\begin{center}
\caption{Same as Table~\ref{table3} except for  $L=0.9\lambdao$.}
\begin{tabular}{||c|c|c|c|c|c||}\hline\hline
 & $n=-2$ &  $n=-1$  & $n=0$ & $n=1$  & $n=2$\\\hline
 $\theta=43^\circ$ & $\mathbf{-1.5402}$ &  $-0.4291$  & $0.6820$ & ${ 1.7931}$ & $2.904$ \\\hline
 $\theta=51.5^\circ$ & $-1.4396$ &  $-0.3285$  & $0.7826$ & $\mathbf{ 1.8937}$ & $3.0048$ \\\hline
 $\theta=55^\circ$ & $\mathbf{-1.4031}$ &  $-0.2920$  & $0.8192$ & ${ 1.9303}$ & $3.0414$ \\\hline\hline
\end{tabular}\label{table10}
\end{center}
\end{table}

 \begin{table}[!ht]
\begin{center}
\caption{Same as Table~\ref{table3} except for  $L=0.75\lambdao$.}
\begin{tabular}{||c|c|c|c|c|c||}\hline\hline
 & $n=-2$ &  $n=-1$  & $n=0$ & $n=1$  & $n=2$\\\hline
 $\theta=22.5^\circ$ &${-2.2840}$ &  $-0.9506$  & $0.3827$ & $\mathbf{ 1.7160}$ & $3.0414$ \\\hline
 $\theta=33.5^\circ$ & $-2.1147$ &  $-0.7814$  & $0.5519$ & $\mathbf{ 1.8885}$ & $3.2186$ \\\hline
 $\theta=46^\circ$ & ${-1.9473}$ &  $-0.6140$  & $0.7193$ & $\mathbf{ 2.0526}$ & $3.3860$ \\\hline\hline
\end{tabular}\label{table11}
\end{center}
\end{table}

In Fig.~\ref{fig15Lap}(a), three peaks are present for all values of $d_1$ at $\theta=43^\circ$, $51.5^\circ$, and $55^\circ$. The relative wavenumbers of several Floquet harmonics at these values of the incidence angle are given in Table~\ref{table10}. The comparison of Tables~\ref{table7} and \ref{table10} shows that each $A_p$-peak represents the excitation of a $p$-polarized SPP wave.  Similarly, each of the three $A_p$-peaks at $\theta=22.5^\circ$, $33.5^\circ$, and $46^\circ$ in Fig.~\ref{fig15Lap}(b) represents the excitation of a $p$-polarized SPP wave as a Floquet harmonic of order $n=1$. The relative wavenumbers of a few Floquet harmonics at these values of the incidence angle are given in Table~\ref{table11}. At each angle, $k_x^{(1)}/\ko$ is close to the real part of one of the $\kappa/\ko$ of the solutions of the canonical boundary-value problem problem provided in Table~\ref{table7}.

\subsection{General conclusions}
In the last two subsections, we have deciphered a host of numerical results and identified those absorbance peaks that indicate the excitation of SPP waves in the grating-coupled configuration, when the dielectric partnering material is periodically nonhomogeneous normal to the mean plane of the surface-relief grating. We found that 
\begin{itemize}
\item[(i)] the periodic nonhomogeneity of the dielectric partnering material enables the excitation of multiple SPP waves of both $p$- and $s$-polarization states;
\item[(ii)] fewer $s$-polarized SPP waves are excited than $p$-polarized SPP waves;
\item[(iii)] for a given period of the surface-relief grating, it is possible for two plane waves with different angles of incidence to excite the same SPP wave (Figs.~\ref{fig1Lap}(a) and \ref{fig15L8});
\item[(iv)] not all SPP waves predicted by the solution of the canonical problem may be excited in the grating-coupled configuration for a given period; 
\item[(v)] the absorbance peaks representing the excitation of $p$-polarized SPP waves are generally wider than those representing $s$-polarized SPP waves; and
\item[(vi)] the absorbance peak is narrower for an SPP wave of higher phase speed (i.e. smaller ${\rm Re}\left[\kappa\right]$).
\end{itemize}
Let us note that some other combination of the periodic functions $\epsilon_d(z)$ and $g(x)$ may allow all solutions of the canonical boundary-value problem to be excited in the grating-coupled configuration.

The solution of the canonical boundary-value problem \cite{Faryadjosab} indicates that the period $2\Omega$ of the rugate filter needs to be greater than a certain value in order for more than one SPP waves to be excited, and the excitation of $s$-polarized SPP waves to exist requires an even larger period. However, the number of possible SPP waves increases as the period   increases up to a certain value. We chose $\Omega=\lambdao$ and $\Omega=1.5\lambdao$ to allow the excitation of multiple $p$-polarized SPP waves   for both values, and multiple $s$-polarized SPP waves  for the second value. Our numerical results confirm that the conclusions on the number of SPP waves drawn in the predecessor paper  \cite{Faryadjosab} also
hold for the grating-coupled configuration.

\section{Concluding Remarks}\label{conc}
The excitation of multiple surface-plasmon-polariton~(SPP) waves by a surface-relief grating
formed by a metal and a dielectric material, both of finite thickness, was studied theoretically using the 
rigorous coupled-wave-analysis technique for the practically implementable setup. The presence of an SPP wave was inferred by a peak in the plots of absorbance vs. the angle of incidence $\theta$, provided that the $\theta$-location of the peak tuned out to be independent of the thickness of the partnering dielectric material. If that material is homogeneous, only one $p$-polarized SPP wave, that too of $p$-polarization state, is excited. However, the periodic nonhomogeneity of the partnering dielectric material normal to the mean plane of the surface-relief grating results in the excitation of multiple SPP waves of different polarization states and phase speeds. In general, the absorbance peak is narrower for an $s$-polarized SPP wave than for of a $p$-polarized SPP wave, and the absorbance peak is narrower for an SPP wave of higher phase speed.

Since the electromagnetic field radiated by a line source can be considered as a spectrum of plane waves propagating at all angles~\cite[Sec.~2.2]{Chew}, the grating-coupled configuration discussed in this paper can be used to excite multiple SPP waves simultaneously by a line source. The excitation of multiple SPP waves may be significant for practical applications---for example, to increase the absorption of light in solar cells due to the increased possibility of excitation of SPP waves~\cite{Ferry}. This application is currently under investigation by the authors.   

\acknowledgments 
MF thanks  the Trustees of the Pennsylvania State University for financial assistance during his doctoral studies.
AL thanks the Charles Godfrey Binder Endowment at the Pennsylvania State
University for ongoing support of his research activities.

 \end{document}